\documentclass{webofc}
\usepackage[varg]{txfonts}   
\usepackage{booktabs, ctable}
\graphicspath{{./figs/}}

\begin{document}

\title{Quarkonium production in high energy $pp$ collisions}
%
%

\newcommand{\jx}[1]{{\color{blue}[\textbf{JX:}\,{#1}]}}

\author{
 \firstname{Jiaxing} \lastname{Zhao}\inst{1},  
 \firstname{Pol Bernard} \lastname{Gossiaux}\inst{1},  
 \firstname{Taesoo} \lastname{Song}\inst{2},  
 \firstname{Elena} \lastname{Bratkovskaya}\inst{2,3},  
 \and   
\firstname{J\"org} \lastname{Aichelin}\inst{1}\footnote{speaker}, 
   }
\institute{ 
 SUBATECH, Nantes University, IMT Atlantique, IN2P3/CNRS
4 rue Alfred Kastler, 44307 Nantes cedex 3, France
 \and
GSI Helmholtzzentrum f\"ur Schwerionenforschung GmbH,
  Planckstr. 1, 64291 Darmstadt, Germany
  \and
 Helmholtz Research Academy Hesse for FAIR (HFHF), 
 GSI Helmholtz Center for Heavy Ion Physics, Campus Frankfurt, 60438 Frankfurt, Germany  }      
\abstract{%
We investigate the charmonium and bottomonium production in $pp$ collisions using the Wigner densities formalism. The Wigner density of the quarkonia is approximated by analytical 3-D isotropic harmonic oscillator Wigner densities with the same root-mean-square radius given by the solution of the  Schr\"odinger equation. This approach reproduces quite well the available experimental transverse momentum and rapidity distributions.}
\maketitle
Hidden heavy flavour mesons are a useful tool to study the strongly interacting quark gluon plasma, which is created during high energy heavy-ion collisions. 
Due to the large quark mass $m_Q$, $m_Q\gg \Lambda_{\rm QCD}$, with $\Lambda_{\rm QCD}$ being the QCD cutoff, quarkonium production can be factorized into the production of a heavy quark, $Q\bar Q$,  pair, which can be described by perturbative QCD and a subsequent soft non-perturbative process, which describes the formation of a colorless quarkonium from the $Q\bar Q$ pair.  For the latter part many approaches have been advanced. They include the Color-Evaporation Model 
, the Color-Singlet Model 
and the Color-Octet Model. 
The latter two are encompassed in the NRQCD approach.
For a review we refer to \cite{Andronic:2015wma}.
More recently, the rapidity and $p_T$ distributions of hidden heavy flavour mesons, produced in $pp$ collisions, have also been well reproduced in the Wigner density matrix~formalism \cite{Song:2017phm,Villar:2022sbv,Song:2023zma}. Here we will extend this formalism up to $3S$ and  use EPOS4 to generate the initial heavy quarks. 

The Wigner density formalism is based on the quantal density matrix projection in which the probability that a meson $i$ is produced is given by $P_i= Tr (\rho_i \rho^{(N)})$ with $\rho_i$ being the density matrix of the meson $i$ and $\rho^{(N)}$ the density matrix of the N heavy quarks and antiquarks, produced in a $pp$ collision. A partial Fourier transformation of the density matrices yields then 
\begin{eqnarray}
\frac{dP_i}{{d^3\bf R}{d^3\bf P}}=\sum \int {d^3rd^3p \over (2\pi)^6}W_{i}({\bf r},{\bf p})\prod_{j>2} \int {d^3r_jd^3p_j \over (2\pi)^{3(N-2)}}W^{(N)}({\bf r}_1,{\bf p}_1,{\bf r}_2,{\bf p}_2,...,{\bf r}_N,{\bf p}_N).
\label{eq.projection}
\end{eqnarray}
$W_i$ is the two-body Wigner density of the bound heavy quark pair and $W^{(N)}({\bf r}_1,{\bf p}_1,{\bf r}_2,{\bf p}_2,...,{\bf r}_N,{\bf p}_N)$ is the quantal density matrix in Wigner representation of the ensemble of N heavy quarks  produced in a $pp$ collision. ${\bf r}({\bf R})$ and ${\bf p}({\bf P})$ are the relative (center of mass) coordinate and momentum of the heavy quark and antiquark, which are bound in a quarkonium.  We assume that the unknown quantal $N$-body Wigner density can be replaced by the average of classical phase space distributions, $W^{(N)}\approx \langle W^{(N)}_{\rm classical} \rangle$. 

The classical momentum space distributions of the heavy quarks is provided by EPOS4~\cite{Werner:2023zvo,Werner:2023fne}.  EPOS4, however, provides only the coordinate information of the vertex where the $Q\bar Q$ pair is created. 
For a heavy quark pair, created at the same vertex, we assume that the relative distance between $Q$ and $\bar Q$  in their center-of-mass frame is given by a Gaussian distribution
and the Wigner density for a $Q\bar Q$ pair can be expressed as
\begin{eqnarray}
W^{(2)}({\bf r},{\bf p})\sim r^2\exp \left(-{r^2 \over 2\sigma_{\rm Q\bar Q}^2} \right)f_{\rm Q\bar Q}^{\rm EPOS4}({\bf p}),
\label{eq.w2body}
\end{eqnarray}  
where the distance is controlled by the effective width $\sigma_{\rm Q\bar Q}$.  Having now momenta and positions of the heavy quarks we can calculate the yield of charmonium and bottomonium via Eq.~\eqref{eq.projection}.
\begin{table}
	\renewcommand\arraystretch{1.0}
	\setlength{\tabcolsep}{1mm}
	\begin{tabular}{c||c|c|c||c|c|c|c|c|c}
		\toprule[1pt]\toprule[1pt]
		\multicolumn{1}{c||}{} & \multicolumn{1}{c|}{$J/\psi$} &   \multicolumn{1}{c|}{$\chi_c(1P)$} &   \multicolumn{1}{c||}{$\psi(2S)$} & \multicolumn{1}{c|}{$\Upsilon(1S)$}& \multicolumn{1}{c|}{$\chi_b(1P)$}& \multicolumn{1}{c|}{$\chi_b(1D)$}& \multicolumn{1}{c|}{$\Upsilon(2S)$}& \multicolumn{1}{c|}{$\chi_b(2P)$}& \multicolumn{1}{c}{$\Upsilon(3S)$}\tabularnewline
		\midrule[1pt]
		Mass Theo.(GeV) & 3.071 & 3.483 & 3.652 & 9.390 & 9.870 & 10.109 & 9.959 & 10.208 & 10.288 \tabularnewline
		\midrule[1pt]
		Mass Exp. (GeV) & 3.097 & 3.463 & 3.686 & 9.460 & 9.876 & 10.163 & 10.023 & 10.243 & 10.355 \tabularnewline		
		\midrule[1pt]
		$\langle r^2 \rangle$ (fm$^2)$ & 0.182 &  0.453 & 0.714 & 0.042 & 0.153 & 0.284 & 0.236& 0.410 & 0.520 \tabularnewline
		\midrule[1pt]
		$\sigma (\rm fm)$ & 0.348 & 0.426 & 0.452&  0.167 & 0.247 & 0.285 & 0.260& 0.302 & 0.307  \tabularnewline
		\bottomrule[1pt]
	\end{tabular}
	\caption{The masses, root-mean-square radius, and the Gaussian widthes $\sigma$ of different charmonium and bottomonium states in vacuum. The experimental data are from Ref.~\cite{Workman:2022ynf}.}
	\label{table1}
\end{table}
\begin{figure}[!htb]
$$\includegraphics[width=0.3\textwidth]{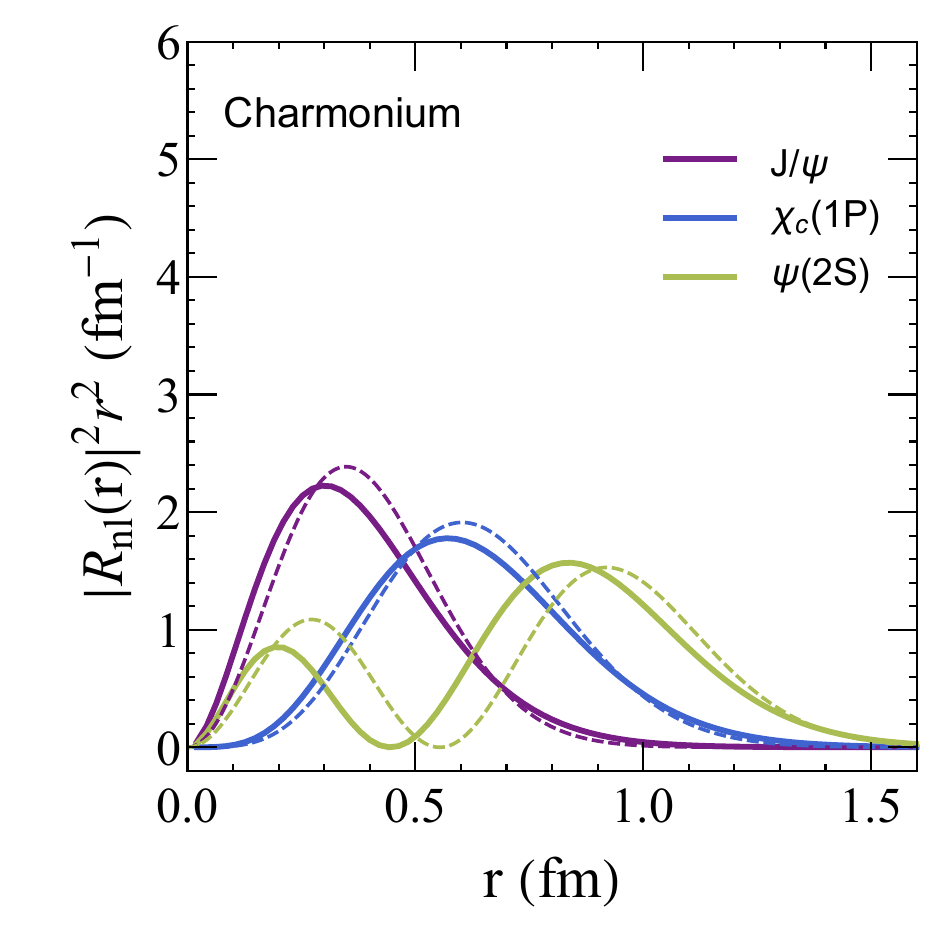}\includegraphics[width=0.3\textwidth]{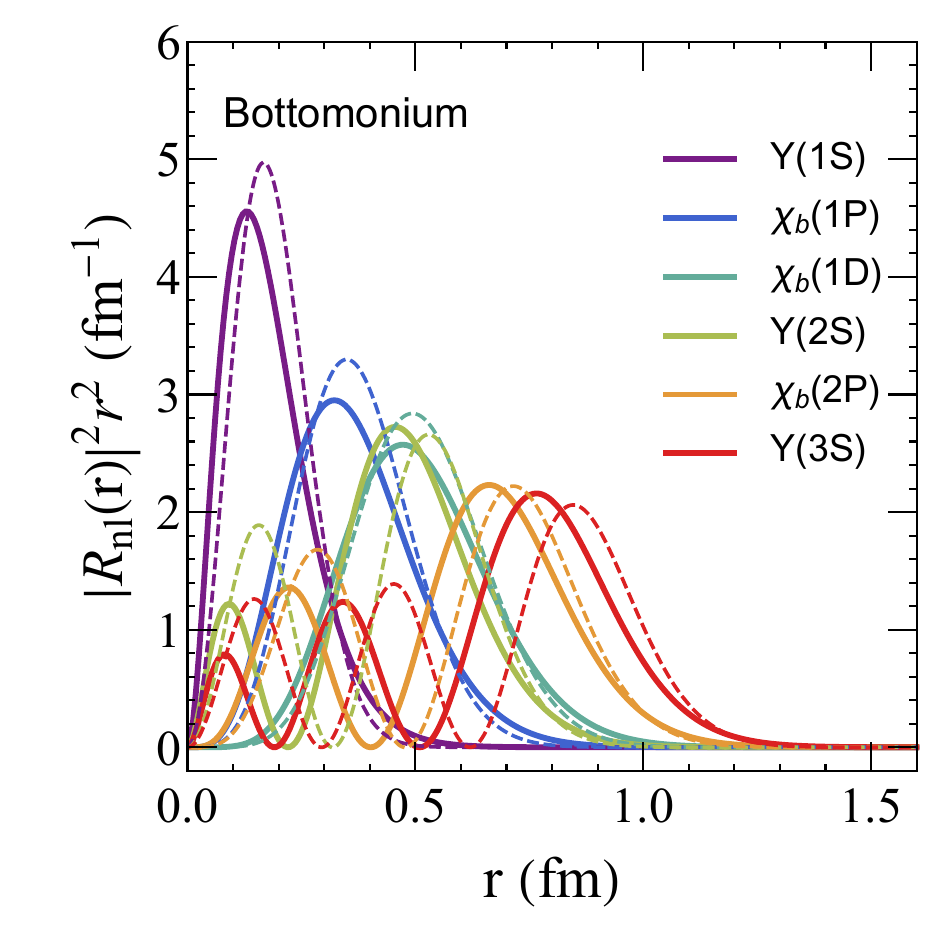}$$
\caption{Wave function of different charmonium (left) and bottomonium (right) states. Solid lines are from the Schr\"odinger equation, dashed lines are from the 3-D isotropic harmonic oscillator (Eq.~\eqref{eq.3dwf}).}
\label{fig.wf}
\end{figure}

We come now to the construction of  the quarkonium Wigner density. The Wigner density is obtained by the Wigner-Weyl transformation of the density matrix of the quarkonia. 
The quarkonium wave function is the  solution of the two-body Schr\"odinger equation, which we solve for charmonium and bottomonium with the Cornell potential, $V(r)=-\alpha/r+\kappa r+c$ with $\alpha=0.513$, $\kappa=0.17 \rm GeV^2$, $c=-0.161$, and with the quark masses $m_c=1.5 \rm GeV$ and $m_b=5.2\rm GeV$. The wave functions are shown in Fig.~\ref{fig.wf}, and the masses and root-mean-square radii $\langle r^2\rangle$ are shown in Table~\ref{table1}. We can see that the masses are very close to the experimental values. 
The wave function is, however,  not analytical, and therefore the Wigner density can only be calculated numerically , what makes the solution of Eq. \eqref{eq.projection}  complicated. It is therefore convenient to approximate the wave function by a 3-D isotropic harmonic oscillator wave function for the potential $V(r)=1/(2m_Q\sigma^4)r^2$. These wave functions are analytical and can be expressed as $\psi_{nlm}(r,\theta,\phi)=R_{nl}(r)Y_{l,m}(\theta,\phi)$, where $Y_{l,m}$ are the spherical harmonics. 
The radial part can be expressed as,
\begin{eqnarray}
R_{nl}(r)=\left [{2 (n!) \over \sigma^3 \Gamma(n+l+3/2)} \right]^{1\over 2}\left({r\over \sigma} \right)^l e^{-{r^2\over 2\sigma^2}}L_n^{l+1/2}\left({r^2\over \sigma^2}\right),
\label{eq.3dwf}
\end{eqnarray}
where $L_n^{l+1/2}$ are Laguerre polynomials. The parameters of  the 3-D isotropic harmonic oscillator wave functions are chosen to match the root-mean-square radius $\langle r^2\rangle$ of the real quarkonium wave function:  $\langle r^2\rangle=3\sigma^2/2$ for $1S$, $\langle r^2\rangle=5\sigma^2/2$ for $1P$, $\langle r^2\rangle=7\sigma^2/2$ for $1D$ and $2S$,  $\langle r^2\rangle=9\sigma^2/2$ for $2P$, and $\langle r^2\rangle=11\sigma^2/2$ for $3S$ states.   The corresponding  widths are shown in Table~\ref{table1} and the wave functions are shown in Fig.~\ref{fig.wf} with dashed lines. We can see that the ground states and low lying excited states can be well reproduced by the 3-D isotropic harmonic oscillator, while the difference increases for higher  excited states, e.g. $2S$, $2P$, and $3S$.

The Wigner densities for different states up to $3S$ are 
\begin{eqnarray}
W_{\rm 1S}({\bf r,p})&=&8e^{-\xi} , \\
W_{\rm 1P}({\bf r,p})&=& {8\over 3}e^{-\xi }\Big(2\xi -3 \Big),\nonumber\\
W_{\rm 1D}({\bf r,p})&=& {8\over 15}e^{-\xi}\Big(15+4\xi^2-20\xi +8[p^2r^2-({\bf p}\cdot{\bf r})^2] \Big),\nonumber\\
W_{\rm 2S}({\bf r,p})&=& {8\over 3}e^{-\xi}\Big(3+2\xi^2-4\xi-8[p^2r^2-({\bf p}\cdot{\bf r})^2)] \Big),\nonumber\\
W_{\rm 2P}({\bf r,p})&=& {8\over 15}e^{-\xi} \Big(-15+4\xi^3-22\xi^2+30\xi-8(2\xi-7)[p^2r^2-({\bf p}\cdot{\bf r})^2] \Big),\nonumber\\
W_{\rm 3S}({\bf r,p})&=& {8\over 315}e^{-\xi} \Big(315+42\xi^4-336\xi^3+924\xi^2-840\xi-[2009+32p^2r^2 \nonumber\\
&+&336r^4/\sigma^4-1400r^2/\sigma^2-896p^2\sigma^2+224p^4\sigma^4][p^2r^2-({\bf p}\cdot{\bf r})^2]-[686\nonumber\\
&+&608p^2r^2+112r^2/\sigma^2-896p^2\sigma^2+224p^4\sigma^4-672({\bf p}\cdot{\bf r})^2]({\bf p}\cdot{\bf r})^2\Big),\nonumber
\end{eqnarray}
where $\xi={r^2\over \sigma^2}+p^2 \sigma^2$.
\begin{figure}[!htb]
\centering
\includegraphics[width=0.3\textwidth]{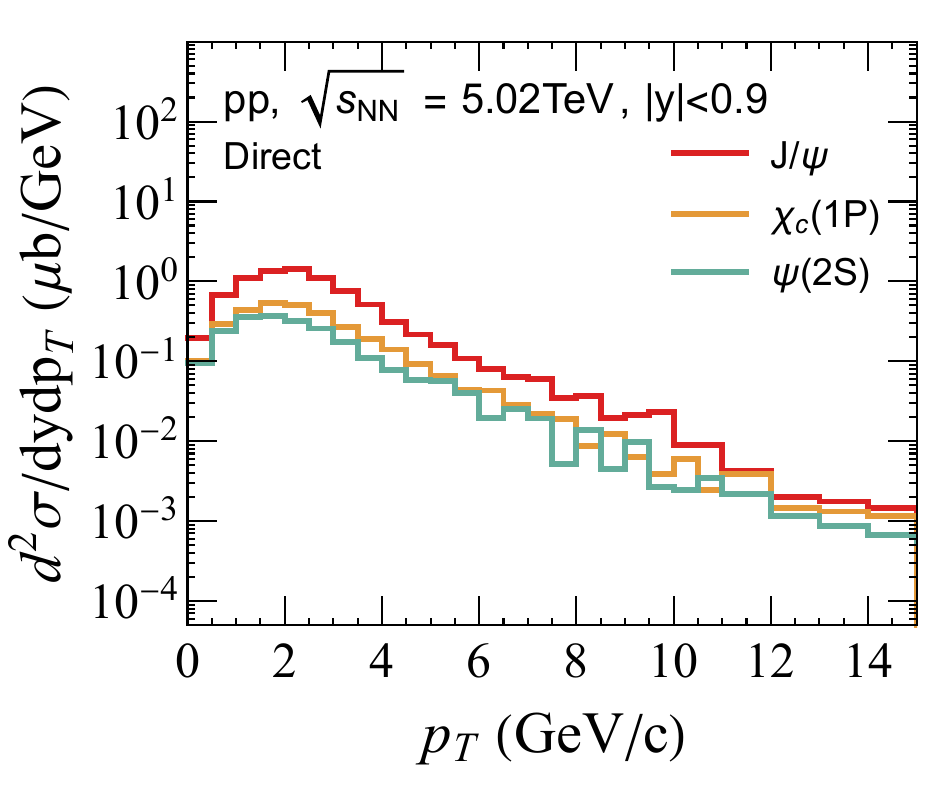}\includegraphics[width=0.3\textwidth]{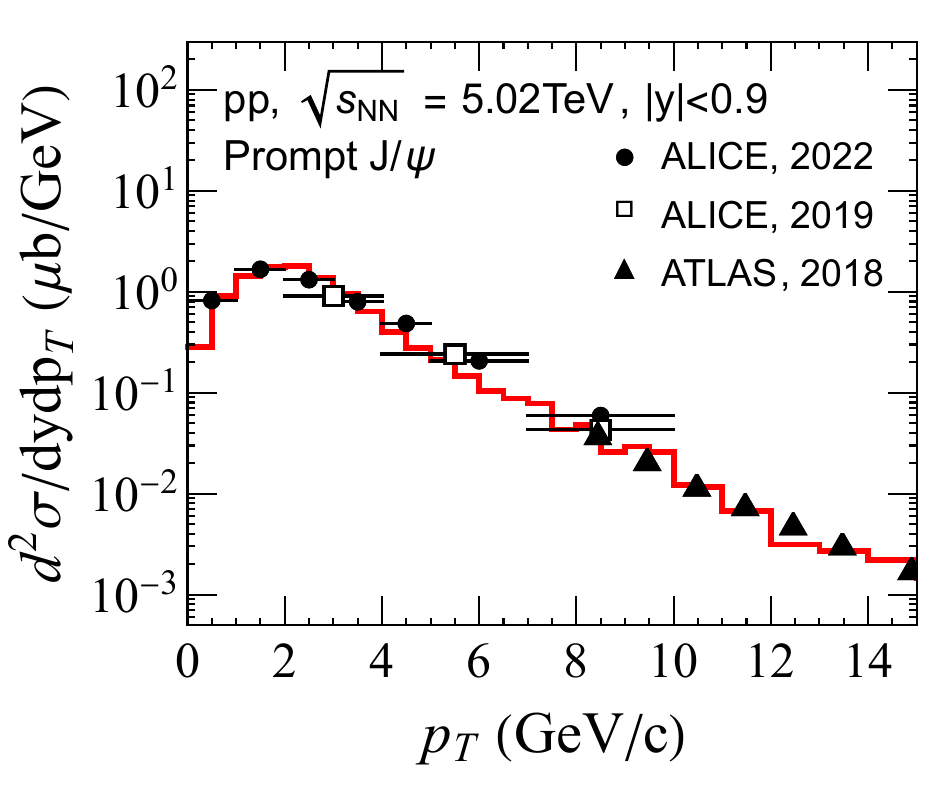}\includegraphics[width=0.3\textwidth]{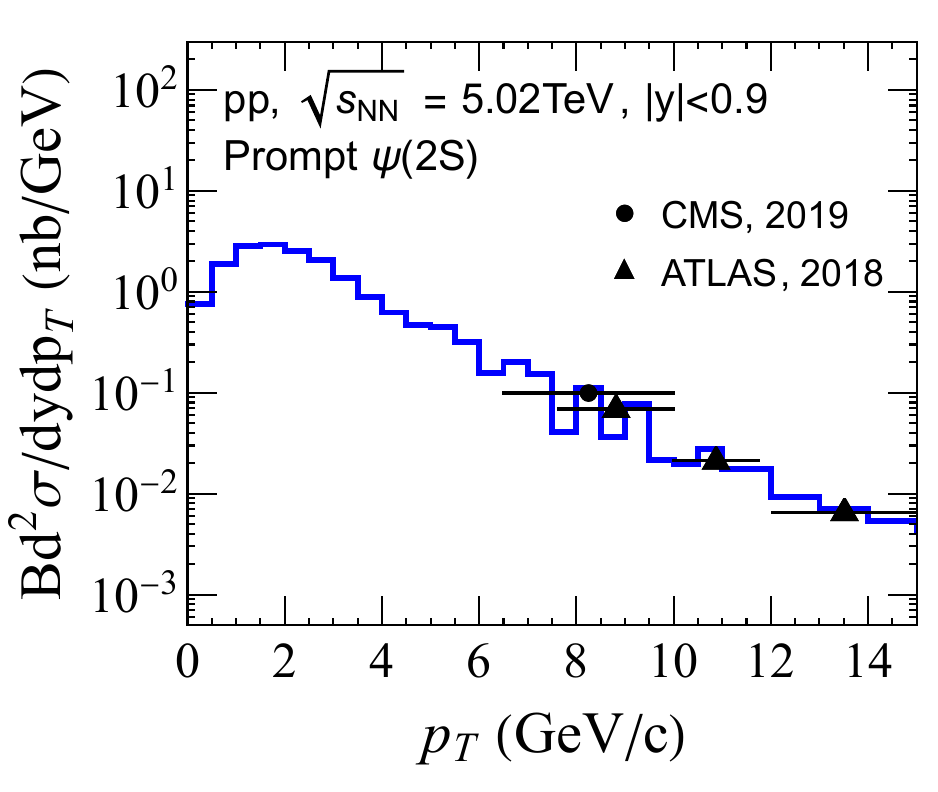}
\caption{$p_T$ spectra of different charmonium states (left), where red is $J/\psi$, orange is $\chi_c(1P)$, and green is $\psi(2S)$. Prompt $J/\psi$ (middle). Prompt $\psi(2S)$ (right). The experimental data are from ALICE~\cite{ALICE:2021edd,ALICE:2019pid}, ATLAS~\cite{ATLAS:2017prf}, and CMS~\cite{CMS:2018gbb}.}
\label{fig.cc}
\end{figure}
\begin{figure}[!htb]
\centering
\includegraphics[width=0.3\textwidth]{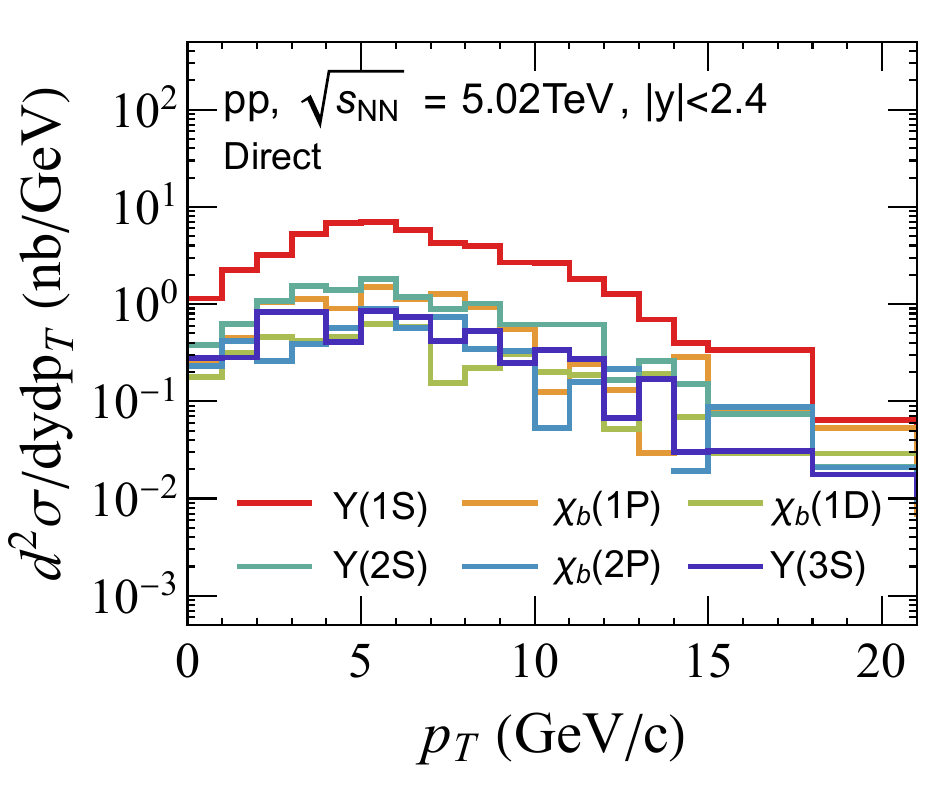}\includegraphics[width=0.3\textwidth]{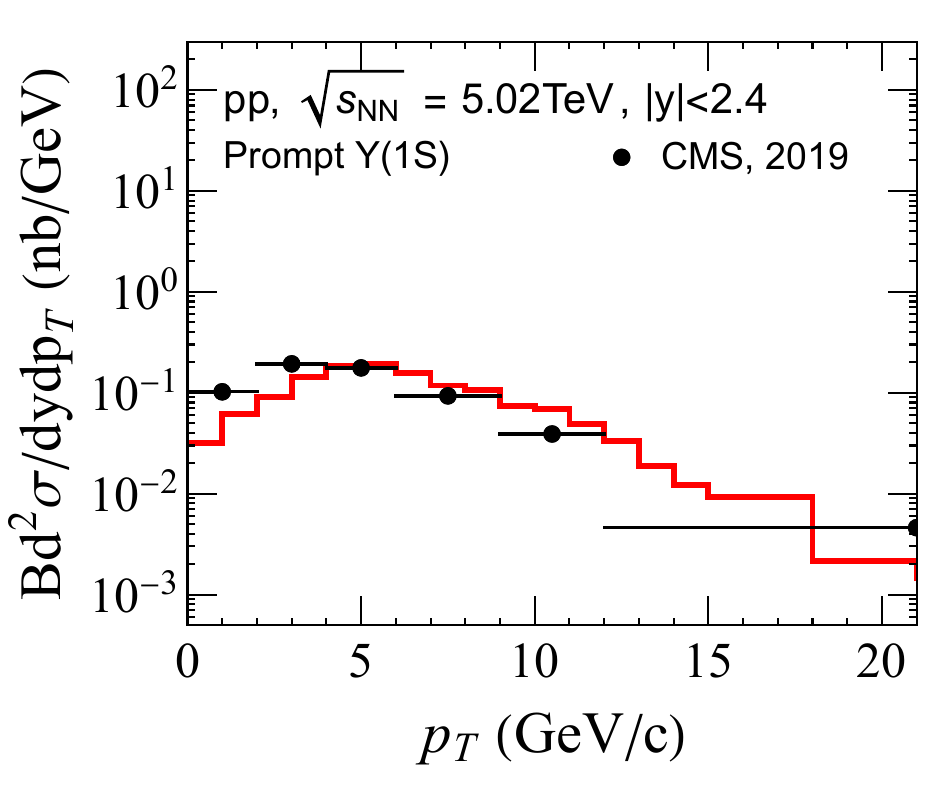}\includegraphics[width=0.3\textwidth]{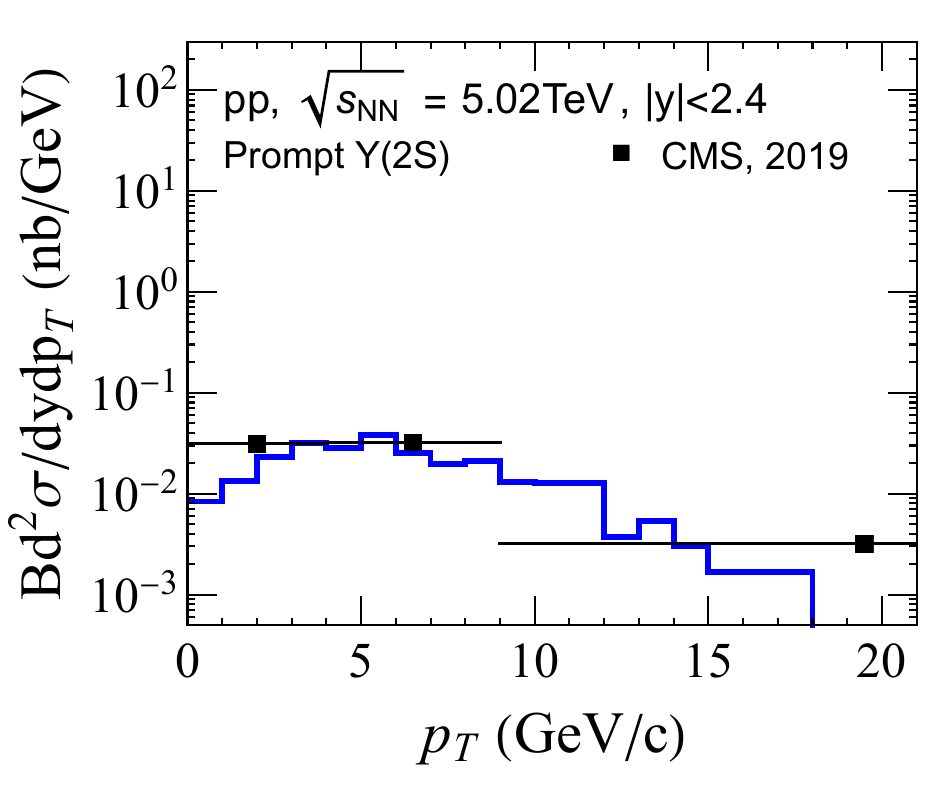}\\
\includegraphics[width=0.3\textwidth]{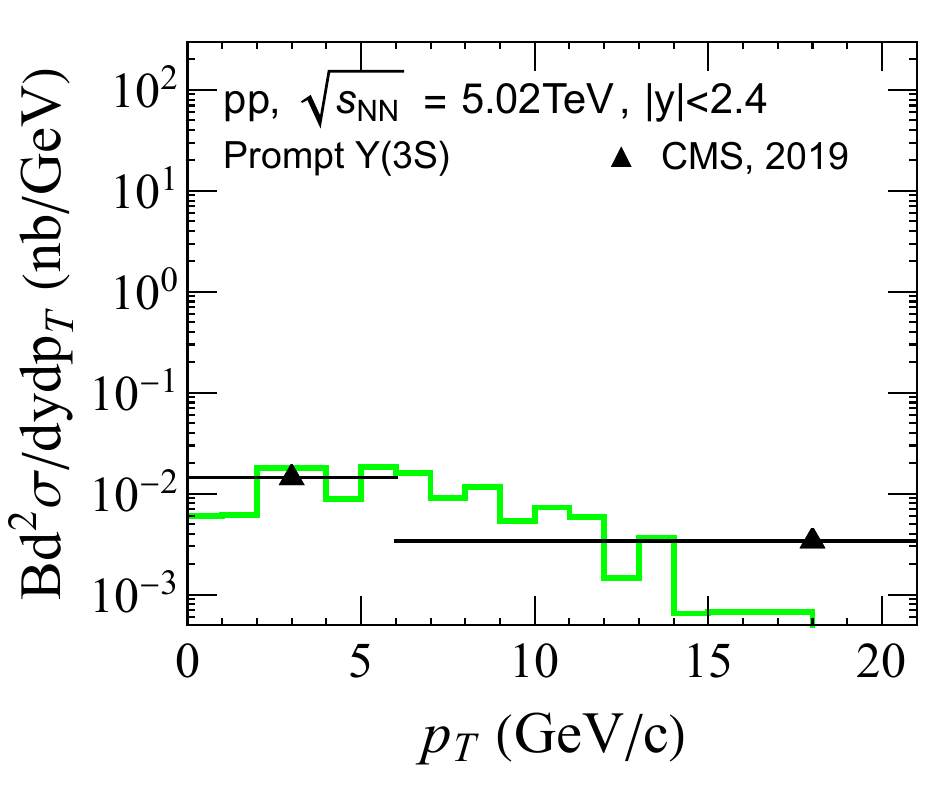}\includegraphics[width=0.3\textwidth]{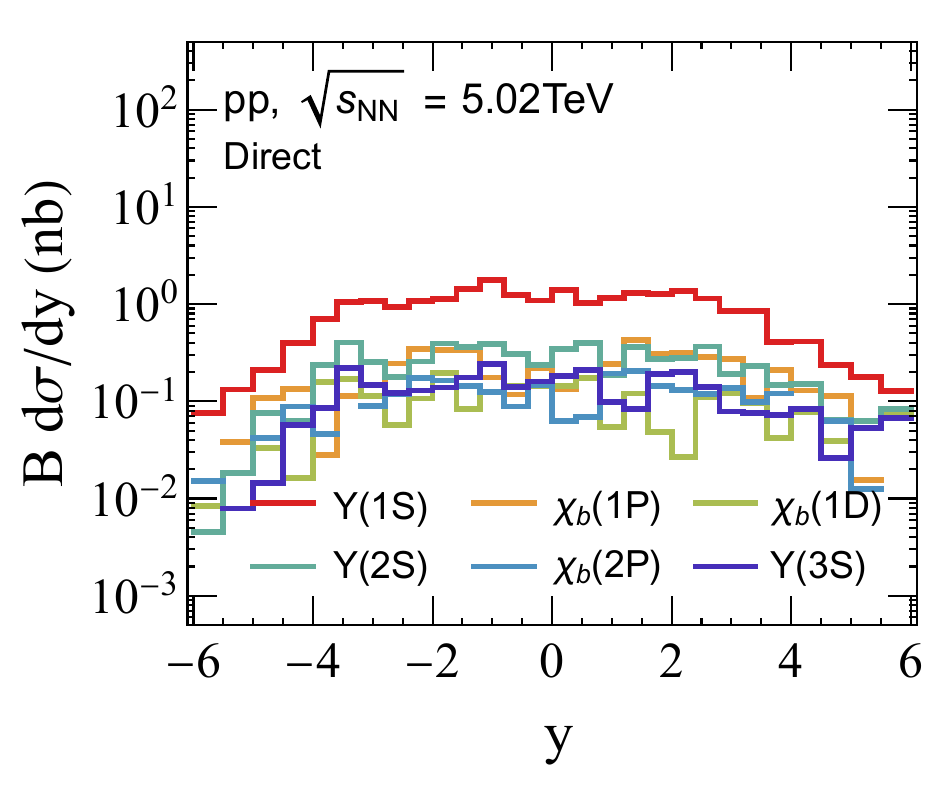}\includegraphics[width=0.3\textwidth]{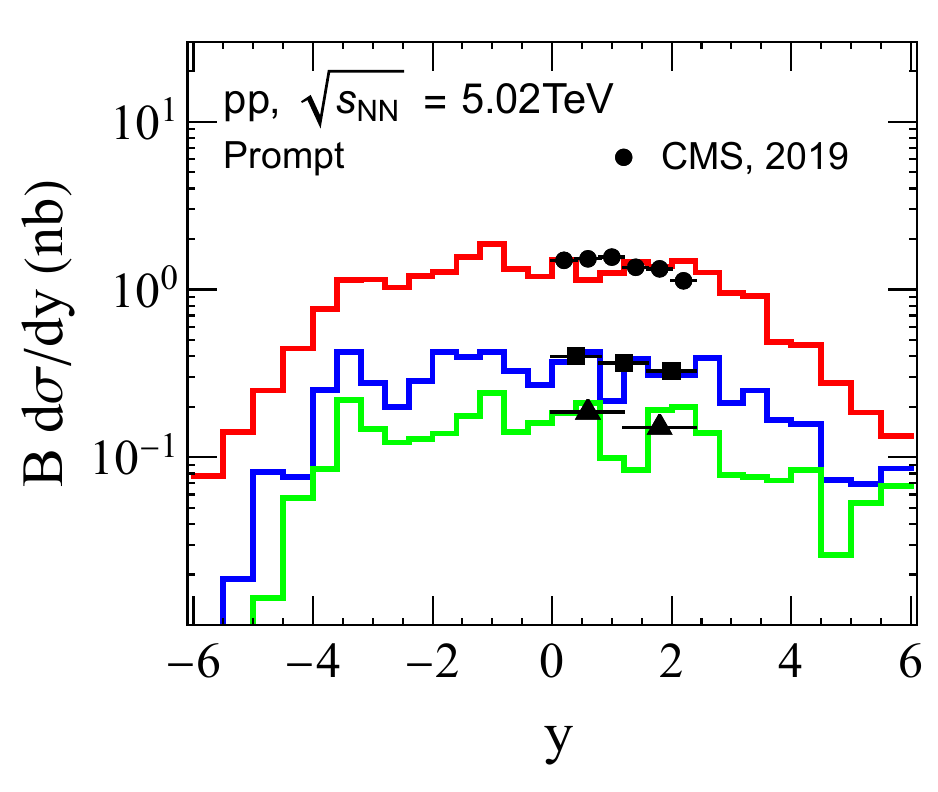}
\caption{$p_T$ and rapidity $y$ dependence of different bottomonium states. The experimental data are from CMS~\cite{CMS:2018zza}.}
\label{fig.bb}
\end{figure}
The transverse momentum and rapidity distribution of charmonium and bottomonium (see Eq.~\eqref{eq.projection}) are shown in Figs.~\ref{fig.cc} and \ref{fig.bb}. We see, as far as data are available, a quite good
agreement with the experimental results for $c\bar c$ as well $b\bar b$ mesons when choosing $\sigma_{c\bar c}$ = 0.4 fm and  $\sigma_{b\bar b}$ = 0.2 fm in Eq.~\eqref{eq.w2body}.

We can conclude that the experimentally available rapidity and transverse momentum distribution of $c\bar c$ and $b \bar b$ quarkonia can be well described in the Wigner density formalism. The only parameter which enters the calculation is the width of the distribution of the relative distance of the $Q\bar Q$ pair at production.  The relative contribution of the different states is then exclusively given by their wave function.  



\vspace*{1mm}
\textit{Acknowledgements:} This work is funded by the European Union’s Horizon 2020 research and innovation program under grant agreement No. 824093 (STRONG-2020). T.S. and E.B. acknowledge support by the Deutsche Forschungsgemeinschaft through the grant CRC-TR 211, Project N 315477589 - TRR 211.

\bibliography{QM2023}

\end{document}